\documentclass [11pt,twoside]{article}

\usepackage{shrthnds}
\usepackage{cite}
\usepackage{graphicx}
\usepackage{color}
\usepackage{subfigure}
\usepackage{setspace}

\usepackage{multirow}

\usepackage[hang,small]{caption}

\usepackage{geometry}
\usepackage{fancyhdr}
\usepackage{titlesec,titletoc}
  \titleformat{\section}{\Large\sf\bfseries}{\thesection}{1em}{}
  \titleformat{\subsection}{\large\sf\bfseries}{\thesubsection}{1em}{}

\title{\sf\bfseries Quantum Treatment of the Weyl Vector Meson}

\author{\normalsize  
Naveen K. Singh$^1$\footnote{email: naveenks@iitk.ac.in}~,
Pankaj Jain$^{1}$\footnote{email: pkjain@iitk.ac.in}~,\\
 Subhadip Mitra$^{2}$\footnote{email: smitra@iopb.res.in} and Sukanta
Panda$^3$\footnote{email: sukanta@iiserbhopal.ac.in}}

\date{}%{\today}

\begin{document}
\maketitle
\vspace{-0.6cm}
\bc
{\small 1) Department of Physics, IIT Kanpur, Kanpur 208 016, India\\
2) Institute of Physics, Bhubaneswar 751 005, India\\
3) Indian Institute of Science Education and Research, Bhopal 462 023, India}
\ec

\centerline{\small\date{\today}}
\vspace{0.5cm}

\bc
\begin{minipage}{0.9\textwidth}\begin{spacing}{1}{\small {\sf\bfseries Abstract:}
%%%%%%%%%%%%%%%%%%%%%%%%%%%%%%%%%%%%%%%%%%%%%%%%%%%%%%
% ABSTRACT
%%%%%%%%%%%%%%%%%%%%%%%%%%%%%%%%%%%%%%%%%%%%%%%%%%%%%%
The Weyl meson arises in theories with local scale invariance. 
It acts as a candidate for dark matter in a generalized Standard Model
with local scale invariance. The Higgs particle is absent from 
the physical spectrum in this theory. We consider the quantization
of this theory in detail, imposing suitable gauge fixing conditions. 
We also consider a further generalization of this model which includes
an additional real scalar field. In this theory a Higgs like particle
remains in the particle spectrum. Since this particle couples to the
Weyl meson, it can lead to interesting phenomenology involving this 
vector field in particle colliders.
%%%%%%%%%%%%%%%%%%%%%%%%%%%%%%%%%%%%%%%%%%%%%%%%%%%%%%
}\end{spacing}\end{minipage}\ec

\vspace{0.5cm}\begin{spacing}{1.1}

\thispagestyle{empty}

%%%%%%%%%%%%%%%%%%%%%%%%%%%%%%%%%%%%%%%%%%%%%%%%%%%%%%
% MAIN CONTENT STARTS HERE
%%%%%%%%%%%%%%%%%%%%%%%%%%%%%%%%%%%%%%%%%%%%%%%%%%%%%%
\section{Introduction}
In recent papers  \cite{JMS,AJS,JM09,JM10_1,JM10_2} we have studied a scale invariant generalization of the
Standard Model of particle physics, including gravity \cite{ChengPRL}.
The model contains no dimensionful parameter and displays local scale invariance
which requires the introduction of the Weyl vector meson.
Local scale invariance was first proposed in Ref. \cite{Weyl:1929}. It was
later revived by many authors
\cite{Dirac1973,Sen:1971,Utiyama:1973,Utiyama:1974,Freund:1974,Hayashi:1976,Hayashi:1978,Nishioka:1985,Ranganathan:1987}.
 Subsequently local scale
invariance has attracted considerable attention in literature \cite{Padmanabhan85,Mannheim89,Hochberg91,Wood92,Wheele
r98,Feoli98,Pawlowski99,Nishino2009,Demir2004,Mannheim09,Moon}.
It has been shown that a scale invariant model of gravity is equivalent to
the Einstein's action with an effective gravitational constant \cite{Deser}.
%The model considered in Refs. \cite{JMS,AJS,JM09,JM10_1,JM10_2}
%contains no dimensionful parameter.
In the model considered in Refs. \cite{JMS,AJS,JM09,JM10_1,JM10_2}, scale
invariance is broken by a soft mechanism analogous to spontaneous symmetry 
breaking \cite{JM,JMS}. Here, motivated by the Big Bang cosmology, we assume a
spatially flat background Friedmann-Robertson-Walker (FRW) metric. The model admits a classical time dependent
solution which breaks scale invariance. This solution is assumed to 
represent the cosmic time evolution.
This generates all the dimensionful
parameters in the theory such as the Hubble parameter, the Planck mass,
the vacuum or dark energy, Weyl meson mass as well as the electroweak
particle masses \cite{JMS,JM09,JM10_1,JM10_2}.
Phenomenological implications
of the Weyl vector meson have been studied in Refs.
\cite{Huang1989,Wei2006,JMS,AJS}.

Scale invariance is generally believed to be anomalous.
Within dimensional regularization,
the scale anomaly arises primarily since the action is not invariant
under scale transformations in dimensions other than four \cite{Delbourgo}.
However it has been shown that any scale transformation can be
extended to arbitrary dimensions such that the action is invariant
under this generalized transformation \cite{Englert:1976,JMS,JM09,JM10_1,JM10_2,Shaposhnikov:2008a,Shaposhnikov:2008b}.
This is accomplished by introducing fractional powers of the scalar 
fields in the action in dimensions different from four.
Hence in the present case, where the
action is exactly invariant in arbitrary dimensions, 
we do not expect
scale invariance to be anomalous. Once the action is specified,
one can extract
dimensionally regularized Feynman amplitudes
directly in $d$ dimensions \cite{Ramond}.

The action displays scale invariance in arbitrary dimensions due to 
introduction of terms with fractional powers of scalar fields. This theory,
therefore, makes sense only if the corresponding field, which is raised
to a fractional power, is non-zero classically. In this case one may
make an expansion around this classical value and the theory is well
defined. Hence we need to demand that scale invariance is broken by some
mechanism, such as spontaneous symmetry breaking, so that the scalar fields
may acquire non-zero value by their classical equations of motion.

The standard model of particle physics generically gives rise to very large
vacuum energy and hence suggests the presence of cosmological constant many
orders of magnitude larger than observations. This has to be cancelled
by explicitly introducing a large cosmological constant term of opposite
sign, leading to a fine tuning problem   
\cite{Weinberg}.
One of our motivation for considering exact scale invariance is that it
is likely to impose some constraints on the cosmological constant
and hence might solve this fine tuning problem. Due
to scale invariance, one is not allowed to introduce a cosmological
constant term in the action \cite{JMS}. 
Furthermore, it has been shown in Ref. \cite{JM10_2}
that pure gauge fields lead to vanishing cosmological constant at
all orders in the gauge field coupling. For the model presented in Ref.
\cite{ChengPRL}, which contains no scalar fields besides the Higgs multiplet,
this follows for the particular
choice of regularization used in Ref. \cite{JM10_2} and applies for all the gauge
fields which do not have any coupling to the Higgs particle. Similar
result applies to fermion fields, as long as we ignore their coupling
to Higgs. However, an effective cosmological constant
is generated in this theory due to the soft 
breakdown of scale invariance. Its value
may be adjusted by fixing the unknown couplings to match the observed dark
energy. The theory also predicts dark matter in the form of the Weyl meson,
which is found to decouple from all the visible matter fields. The model,
therefore, correctly describes all the particle and cosmological data.

The model does introduce some
parameters which take very small values. This is basically associated with
the smallness of the cosmological constant. Although many contributions
to vacuum energy vanish due to scale invariance, it may still acquire
large non-zero values at loop orders. One may suitably adjust the model
so that matter sector gives zero contribution to the cosmological constant \cite{JMPS}. 
However in this case one is still required to fine tune the potential
to set some parameters to zero. In the present paper we shall not address
the problem of fine tuning of the cosmological constant.

In this paper we are interested in identifying the quadratic terms
and the propagators 
of the Weyl meson and the scalar fields with which it couples, 
after introducing suitable 
gauge fixing conditions. We are primarily interested in applications
to processes which involve the Weyl meson. The interactions of the Weyl 
meson are determined by the corresponding gauge coupling which we denote
by $f$. We are interested only in the dominant contributions which
might lead to signals accessible in current or future colliders or in 
cosmology. We are not interested in contributions which are 
proportional to the gravitational coupling since they play a role only
in very early times in cosmology and hence are practically not accessible
to observations.   

This paper is organized as following. In section \ref{sec:realscalar} we discuss a toy model with a single scalar field. We identify the particle content of this model by choosing an appropriate gauge condition. In section \ref{sec:sism} we consider an extension of the Standard Model such that it displays local scale invariance. The only scalar field in this model is a single Higgs multiplet. In this case the Higgs boson gets eliminated from the physical particle spectrum in this model, as it acts essentially like the longitudinal mode of the Weyl meson \cite{ChengPRL}. However, as we shall see, such a theory becomes non-perturbative beyond the electroweak scale. Hence in section \ref{sec:dilaton} we consider a generalization of the model with only one extra scalar field. In this case the Higgs particle is present in the physical spectrum. Finally we conclude in section \ref{sec:conclusion}.

\section{Scale Invariant Model with a Real Scalar Field}\label{sec:realscalar}
%In this section we consider the model with a single real scalar field besides
%gravity and the Weyl meson.
Consider the following action in $d=4-\epsilon$ dimensions
containing only gravity, the Weyl meson $S_\mu$ and a scalar field $\chi$:
\ba
\mathcal{S} &=& \int d^dx \sqrt{-\bar g}\Bigg[{\beta\over 8} \chi^2
\bar R' + \frac12\bar g^{\mu\nu} (D_\mu\chi)(D_\nu\chi)
- {1\over 4}\bar g^{\mu\rho}\bar g^{\nu\sigma}\mathcal{E}_{\mu\nu}
\mathcal{E}_{\rho\sigma}
(\chi^2)^{\delta} %\nn\\&&
 - {1\over 4}\lambda \chi^4 (\chi^2)^{\, -\delta} \Bigg] ,
\label{eq:toy model}
\ea
where $\delta = (d-4)/(d-2)=-\epsilon/(2-\epsilon)$,
$\mc{E}_{\mu\nu}$ represents the field strength tensor for $S_\m$ and $D_\m$ is the scale covariant derivative with
$f$ being the gauge coupling. The symbol
$\bar R'$ represents the scale covariant curvature scalar
\ba
{\bar R'}={\bar R} + 4\frac{d-1}{d-2}f  S_{;\mu}^\mu + 4\frac{d-1}{d-2}f^2  S^{\mu}S_{\mu}.
\label{eq:R'bar}
\ea
We choose the conventions followed in Refs.
\cite{Donoghue1} where
the flat space-time metric takes the form $(1,-1,-1,-1)$.
The curvature tensor and its contractions are defined as
\ba
 R^\mu_{\nu\alpha\beta}&=& -\partial_{\beta}\Gamma^\mu_{\nu\alpha} +\partial_{\alpha}\Gamma^\mu_{\nu\beta} +\Gamma^\mu_{\gamma\alpha}\Gamma^\gamma_{\nu\beta}
 -\Gamma^\mu_{\gamma\beta}\Gamma^\gamma_{\nu\alpha},\nn\\
R_{\nu\beta} &=& R^\mu_{\nu\beta\mu} \hspace{1mm} ,\hspace{4mm} R = R_{\nu\beta}g^{\nu\beta}.
\label{eq:R_notation}
\ea
In Eq. \ref{eq:toy model}, $\beta$ and $\lambda$ denote the
coupling parameters. In Eq. \ref{eq:toy model} we also have terms with field
$\chi$ raised to fractional powers if the space-time dimension is different
from four. These terms are necessary in order to maintain scale invariance
at the quantum level \cite{Englert:1976,JMS,JM09,Shaposhnikov:2008a,Shaposhnikov:2008b}. This is useful since we are demanding local scale invariance.
In this case there would be no anomalous contributions to the
Ward identities which may be useful in maintaining renormalizability of
the model. We point out that the full model is not renormalizable due to
quantum gravity contributions, in any case. However we expect that at least
up to a mass scale much larger than the electroweak scale, such problems
may be ignorable \cite{Shaposhnikov09}. We also point out that all the
predictions of the broken scale invariance are maintained by this procedure
\cite{Shaposhnikov:2008a,Shaposhnikov:2008b}. Hence we still find scale 
dependence of the coupling constants exactly as expected due to anomalous
contributions. However now these contributions arise due to the fractional
powers of the fields in dimensions other than four and due to the fact that
 a scale is already
present in the theory due to a soft breakdown of scale invariance.

We next analyse this model in four dimensions. We seek a classical solution
with $S_\mu=0$ and constant $\chi$. The equation of motion for
the scalar field gives
\ba
{\beta R\over 4} &=& \lambda\chi^2\,.
\label{eq:auxi}
\ea
Here $R$ represents the curvature scalar, as
defined in Eq. \ref{eq:R_notation}. 
We have also set the derivative terms of the scalar field equal to zero. 
We obtain a solution to this classical equation such that the
scalar field, $\chi=\chi_0$, is constant and 
the FRW scale parameter is given by,
\ba
a(t) = a_0e^{H_0t},
\label{eq:at}
\ea
where $H_0$ is the Hubble parameter. This sets the curvature scalar 
as $R=12H_0^2$, where
we have taken the FRW curvature parameter $k=0$. Hence we obtain a
de Sitter space-time. In Ref. \cite{Purnendu} it has been shown
that this classical solution is stable under small perturbations. 
We assume this solution approximately represents the
cosmic evolution today which is dominated by dark energy. We may obtain
a solution closer to observations by including dark matter. Remarkably
in the present model this is provided by the Weyl meson \cite{ChengPRL,AJS}.
In the present paper we ignore these refinements and assume that the
background is described by pure de Sitter metric.
The constant $\chi_0$ is related to the  Planck mass, $M_{\rm PL}$, by,
\ba
\chi = \chi_0 = {M_{\rm PL}\over \sqrt{2\pi\beta}}\,,
\label{eq:chi_Mpl}
\ea
so that classically we obtain the observed gravitational constant.
%The scalar curvature given by
%\ba
%R= 4\lambda\chi_0^2/\beta\, .
%\label{eq:classical_R}
%\ea

Under a global scale or global 
pseudoscale transformation \cite{ChengPRL,JMS}, both $R$
and $\chi$ transform so as to leave Eq. \ref{eq:auxi} invariant.
However, since $M_{\rm PL}$ remains unchanged, the solution breaks
scale invariance. 
Hence the symmetry is broken by the classical time dependent
solution which represents the background cosmic evolution. 
The phenomenon is not the same as spontaneous symmetry breaking 
or dynamical symmetry breaking. In both of these cases the ground
state is not invariant under the corresponding symmetry transformation
\cite{IZ}. 
In the present case the symmetry is broken by a classical time 
dependent solution. The corresponding quantum state may be obtained
by making a small perturbation around this solution and quantizing
the fluctuation modes. This is similar to the quantization of solitons
where one also makes a quantum expansion around a classical solution which
may depend on space and/or time \cite{Rajaraman}. The procedure is 
straightforward as long as one considers only the matter fields \cite{JM}. 
In this case it is clear that as we quantize this theory in the background
of a classical time dependent solution the lowest
energy state is not the true ground state of the theory \cite{JM}. Due 
to time dependence it will clearly have energy higher than the ground
state.  The present case is complicated due to the
usual problems associated with quantization
of gravity. Even the basic interpretation of the wave function of the 
universe is unclear. In any case even in the presence of gravity
the symmetry breaking mechanism being considered here is very different
from the standard spontaneous symmetry breaking since it is being generated
by a background time dependent solution rather than the minimum of the
potential. 
Solutions similar to Eq. \ref{eq:auxi} have been considered earlier in Refs.
\cite{Shore,Cooper,Allen,Buchbinder85,Finelli}.
From Eq. \ref{eq:chi_Mpl} we see that the solution correctly generates the
gravitational constant.

The value of $R$ is fixed by the observed value of the Hubble constant.
Classically,
the parameter $\beta$ is fixed by Eq. \ref{eq:chi_Mpl} and Eq. \ref{eq:auxi}
fixes the
value of the parameter $\lambda$. This parameter essentially fixes the
cosmological constant in this model by the relationship
\ba
\rho_V = {1\over 4} \lambda \chi_0^4.
\label{eq:rhoV}
\ea
The parameter $\lambda$
takes a very small value in this model and so it is important to check
if there are any loop corrections which tend to drive it to very large
values. There may be large quantum corrections to the cosmological
constant. Here we shall absorb these corrections in the parameter 
$\lambda$ and may lead to fine tuning of this parameter.
So far the fine tuning problem of this parameter is unsolved. 
Hence we find that the well known fine tuning problem of the 
cosmological constant appears in our model 
as a fine tuning problem of the parameter $\lambda$. 

We next point out that, in vacuum, if we assume spherical symmetry we
reproduce the standard Schwarzschild-de Sitter or Kottler space-time solution
\cite{Kottler,Rindler} to standard Einstein's gravity with a cosmological constant.
We impose the boundary conditions on the scalar field such that at large 
distances it approaches a constant value and its first derivative approaches 
zero. We find that the solution which satisfies the equations and
these boundary conditions is 
\be
ds^2 = f(r) dt^2 - {dr^2\over f(r)} - r^2
\left(d\theta^2 + \sin^2\theta d\phi^2\right)
\ee  
where
\be
f(r) = 1-{2M\over r} - {1\over 3}\Lambda r^2
\ee
with the scalar field $\chi$ is equal to $\chi_0$, which is a constant.
Here $M$ is the mass of the source and $\Lambda$ the cosmological constant. 
In our case $\Lambda = \lambda\chi_0^2/\beta$. It is clear that the extra
term proportional to $\Lambda$ in the metric will have negligible 
effect on physics on the scale of the solar system due to its
very small value. This is discussed in detail in Ref. \cite{Kagramanova}. 
Furthermore it has no effect on the bending of light \cite{Lake}. Hence
we find that our model gives predictions in agreement with the standard
Einsteins gravity at distances of the order of the solar system. 

We next make a quantum expansion around this solution with
\ba
\chi = \chi_0 + \hat \chi.
\label{eq:chihat}
\ea
We shall set
\ba
\big<\hat\chi\big> = 0
\label{eq:hatchi}
\ea
as a renormalization condition. Choosing this condition implies that
$\big<\chi\big>=\chi_0$,  
to all orders in perturbation theory. This is 
analogous to the renormalization condition normally imposed on
theories with spontaneous symmetry breaking \cite{Peskin}.  

We also expand the graviton field around its classical solution,
We have \cite{tHooft}
\begin{eqnarray}
\bar g_{\mu\nu} &=& g_{\mu\nu} + h'_{\mu\nu},\cr
\bar g^{\mu\nu} &=& g^{\mu\nu} - h^{\prime\mu\nu} + h_\alpha^{\prime\mu}
h^{\prime\alpha\nu}, \cr
\sqrt{-\bar g} &=& \sqrt{-g}\left(1+{1\over 2}h^{\prime\alpha}_\alpha -
{1\over 4} h^{\prime\alpha}_\beta h^{\prime\beta}_\alpha + {1\over 8} (h^{\prime\alpha}_\alpha)^2\right).
\end{eqnarray}
%We next analyze the quadratic Lagrangian in detail in order to fix the gauge
%and determine the propagators of different particles.
The Ricci scalar $\bar R$ may be expanded as
\begin{equation}
\bar R = R + h^{\prime\beta;\alpha}_{\beta;\alpha}- h^{\prime\beta;\alpha}_{\alpha;\ \beta}
- R^\alpha_\beta h^{\prime\beta}_\alpha + R_2,
\end{equation}
where $R_2$ contains all the second order terms in the graviton field $h_\alpha^\beta$:
\begin{eqnarray}
R_2 &=& -{1\over2}D_{\alpha}\left( h_{\mu}^{'\beta}h_{\beta}^{'\mu ;\alpha}\right) +{1\over2}D_{\beta}\bigg\{h_{\nu}^{'\beta}\left(2 h_{;\alpha}^{'\nu \alpha}-h_{\alpha}^{'\alpha ;\nu}\right)\bigg\} \nonumber \\
&&+ {1\over4}\left( h_{\beta ; \alpha}^{'\nu} +  h_{\alpha ; \beta}^{'\nu} -  h_{\beta \alpha }^{'; \nu}\right)
    \left( h_{\nu}^{'\beta ;\alpha} +  h^{'\beta  \alpha}_{;\nu} -  h_{\nu }^{'\alpha ; \beta }\right)-{1\over4}\left(2 h_{;\alpha}^{'\nu \alpha}-h_{\alpha}^{'\alpha ;\nu}\right)h^{'\beta}_{\beta ;\nu}\nonumber \\
&& - {1\over2}h^{'\nu\alpha}
h_{\beta ;\nu \alpha}^{'\beta }+{1\over 2}h_{\alpha}^{'\nu}D_{\beta}\left( h_{\nu}^{'\beta ;\alpha} +  h^{'\beta  \alpha}_{;\nu} -  h_{\nu }^{'\alpha ; \beta }\right)
+h_{\beta}^{'\nu}h_{\alpha}^{'\beta}R_{\nu}^{\alpha}.
\end{eqnarray}
As mentioned in the introduction, we are not interested in quantum gravity
contributions since these are negligible in phenomenological
and cosmological applications, excluding the very early phase. 
However it is useful to keep the quadratic terms in the
graviton since these will mix with the scalar field and the Weyl meson. 
Hence these are necessary to properly identify the particle spectrum
in the theory. 

To suitably normalize the graviton field such that it
has the properly normalized kinetic energy term, we define
\begin{equation}
h_\alpha^\beta = {1\over 4} \chi_0\sqrt{\beta} h_\alpha^{\prime\beta}.
\label{eq:redefine}
\end{equation}
%We denote the trace $h_\alpha^\alpha$ by the symbol $h$.
Quantum expansion of the Lagrangian gives the following quadratic terms
\ba
 \sqrt{-\bar g} \underline{\underline{\mc L}} &=& \sqrt{-g}\bigg[{1\over 2}f^2
\chi_0^2 \left(1+{3\beta\over 2}\right) S^{\mu}S_{\mu}
+ {1\over 2} D^\mu\hat\chi D_\mu\hat \chi
- {1\over 4}g^{\mu\rho}
g^{\nu\sigma}{\mc E}_{\mu\nu}{\mc E}_{\rho\sigma}  \cr
&& - \lambda\chi_0^2   \hat\chi ^2
+ \sqrt{\beta}\hat\chi \left( h^{\beta;\alpha}_{\beta;\alpha}-  h^{\beta;\alpha}_{\alpha;\ \beta}
- R^\alpha_\beta  h^\beta_\alpha\right)\cr
&&+ {R_2\over 8} \beta \chi_0 ^2
+ h\left( h^{\beta;\alpha}_{\beta;\alpha}-  h^{\beta;\alpha}_{\alpha;\ \beta}
- R^\alpha_\beta  h^\beta_\alpha\right)
+ f S^{\mu}_{; \mu}\left\{\left({3\beta\over 2}+1\right) \chi_0 \hat\chi
+\frac32{\sqrt{\beta} \chi_0  } \ h\right\}\cr
&&+4 {\lambda\chi_0^2\over\beta}\left(-{1\over 4} h^\alpha_\beta
 h^\beta_\alpha + {1\over 8}  h^2\right)
 + \cdots \bigg].
\ea

\subsection{Weyl Meson in the $R_\xi$ Gauge}
We may eliminate the mixing terms of $S_{\mu}$ with other fields by adding
the following gauge fixing term
\ba
{\mc L}_{gf} = -{1\over 2\xi}\left[S^\kappa_{;\kappa}
+\xi \left(C_1\hat\chi + C_2 h \right)\right]^2,
\ea
where
\ba
C_1 = \left({3\beta\over 2}+1\right)f\chi_0;  \ \ C_2 = \frac32\sqrt{\beta}f \chi_0.
\ea
Hence we find
\ba
 \sqrt{-\bar g} \underline{\underline{\mc L}} &=& \sqrt{-g}\bigg[{1\over 2}
D^\mu\hat\chi D_\mu\hat\chi+
\sqrt{\beta} \hat\chi \left( h^{\beta;\alpha}_{\beta;\alpha}-  h^{\beta;\alpha}_{\alpha;\beta}
\right)
- {1\over 4}g^{\mu\rho}
g^{\nu\sigma}{\mc E}_{\mu\nu}{\mc E}_{\rho\sigma}  +  R'_2\cr
&&
- {1\over 2 \xi}\left( S^{\mu}_{;\mu} \right)^2
+ {1\over 2}f^2 \chi_0^2 \left(1+{3\beta\over 2}\right) S^{\mu}S_{\mu}
+   \bigg\{-\chi_0 ^2 \lambda
-{\xi\over 2}C_1^2 \bigg\} \hat\chi ^2 \cr
&&- 4{\lambda\chi_0^2\over \beta}  
\left(-{1\over 4} h^\alpha_\beta
 h^\beta_\alpha + {1\over 8}  h^2\right) - {\xi\over2}C_2^2 \ h^2\cr
&&
-\left({\sqrt{\beta} R \over 4 } +\xi C_1C_2\right)\hat\chi h
+\cdots \bigg],\ea
where $R_2'$ represents the kinetic energy terms of gravitons.

The mass terms of $\hat\chi$ and $h$ may be written as
\ba
{\mc L}_m = -{1\over 2}\left(\xi C_1^2 + \epsilon_1\right)\hat\chi^2
- {1\over 2}\left(\xi C_2^2 + \epsilon_2\right)h^2
- \left(\xi C_1 C_2 + \epsilon_3\right)\hat \chi h,
\ea
where the $\epsilon_i$ are of order $\lambda\chi_0^2$ and hence very small.
The mass matrix may be diagonalized to obtain two particles of masses
\ba
M^2 \approx \{\xi(C_1^2 + C_2^2), 0\}.
\ea
It is clear from the above analysis that the Goldstone like mode $S_{\rm G}$
that gets eliminated to generate the mass of the Weyl meson is the linear combination
\ba
S_{\rm G} = {1\over \sqrt{C_1^2+ C_2^2}}\left(C_1\hat \chi + C_2 h\right)
\label{eq:SG1}
\ea
with $M^2\approx \xi(C_1^2 + C_2^2)$. The propagator of the Weyl meson
may be written as
\begin{equation}
i\Delta^S_{\mu\nu}(k) = {-i\over k^2 - M_S^2 + i\epsilon}\left[g_{\mu\nu}
- (1-{\xi}) {k_\mu k_\nu\over k^2 - \xi M_S^2 } \right],
\label{eq:Weyl_propagator}
\end{equation}
where,
\ba
M_S = f\chi_0\sqrt{1+3\beta/2}
\ea
is the mass of the Weyl meson.

So far we have confined our discussion to dimension $d=4$. In order
to compute quantum corrections we need to regulate the theory.
We use dimensional regularization such that scale invariance is
preserved. As mentioned in the introduction, the scale anomaly arises
due to the fact that the process of regularization breaks scale invariance
\cite{Delbourgo}. 
However, in this case the fractional powers of the scalar field $\chi$ in dimensions different from four helps to preserve scale invariance in the dimensionally regulated the action \cite{Englert:1976,JMS,JM09,Shaposhnikov:2008a,Shaposhnikov:2008b}. 
The field with fractional power can be handled by making a 
quantum expansion around its classical value. For example, the terms such
as $(\chi^2)^\delta$, appearing in Eq. \ref{eq:toy model}, may be expanded
such as,
\ba
(\chi^2)^\delta = \chi_0^{2\delta} \left( 1+2\delta{\hat\chi\over\chi_0}
+ \cdots\right).
\label{eq:regulator}
\ea
by using Eq. \ref{eq:chihat}.
Hence the expansion,
Eq. \ref{eq:regulator}, is an expansion in powers of
$\hat\chi / \chi_0$. The terms in this expansion will
lead to additional Feynman rules involving the scalar field.
It is clear that this expansion makes sense only if the classical value,
$\chi_0$, is non-zero. Hence this mechanism works only if scale invariance
is broken by some soft mechanism.

We stress that this procedure does not conflict with the 
standard predictions of scale anomaly \cite{Shaposhnikov:2008a,Shaposhnikov:2008b}. The scale dependence of the coupling
constant is also observed in these models, despite the fact that scale
invariance is not anomalous. Here the scale dependence arise due to the
fact that a scale is already present in the theory due to a soft breakdown
of scale invariance. 

The scale invariant action in arbitrary dimensions is shown in Eq.
\ref{eq:toy model} for the case of a single scalar field besides
gravity. More fields can be added as discussed in Ref. \cite{JM10_1}.
As we expand around the classical solution, Eq. \ref{eq:regulator},
we shall generate an infinite series of terms involving higher powers
of $\hat\chi/\chi_0$.
It is clear that the contribution of these terms is suppressed
by powers of $p^2/ M_{\rm PL}^2$, where $p$ is the momentum scale of
the process under consideration \cite{Shaposhnikov09}.

\section{Standard Model with Local Scale Invariance}\label{sec:sism}
In this section we consider a scale invariant generalization of the Standard Model,
originally proposed in Ref. \cite{ChengPRL}.
Here the only scalar field present is the Higgs multiplet. The Higgs
field acts as the longitudinal mode of the Weyl meson and hence
decouples from the physical spectrum.
The action for the scale invariant extension
of the Standard Model in
$d=4-\epsilon$ dimensions may be written as \cite{JM10_1}
\ba
\mathcal{S} &=& \int d^dx \sqrt{-\bar g}\Bigg[{\beta\over 4} \mc H^\dag \mc H
\bar R'
+\bar g^{\mu\nu} (D_\mu \mc H)^\dag(D_\nu \mc H)  
 - {1\over 4}\bar g^{\mu\rho}\bar g^{\nu\sigma}\mathcal{E}_{\mu\nu}\mathcal{E}_{\rho\sigma}
(\Phi^2)^{\delta}\nn\\ 
&& -\lambda    (\mc H^\dag \mc H)^2 (\Phi^2)^{\, -\delta}  \Bigg],
\label{eq:S_EW_d}
\ea
where $\mc{H}$ is the Higgs doublet, 
$\mc{E}_{\mu\nu}$
represent the field strength tensor of the 
the Weyl
vector field. Here we have not displayed the remaining vector and
fermionic fields since these do not directly couple to the Weyl meson 
\cite{ChengPRL} and hence play no role in our analysis. 
The symbol, $\Phi$ denotes a scalar, which we
choose to be $\Phi^2=\mc H^\dag \mc H$.

The scale invariance is broken by a classical solution to the equations
of motion. Here we shall consider the specific model with only one scalar
field multiplet, displayed in Eq. \ref{eq:S_EW_d}.
In general, there might be other scalar fields present, in which case
the solution discussed below may be suitably generalized.
%This is discussed in Section 3.
We denote the classical solution to the Higgs field by $\mc H_0$. We find a
solution to the classical equations of motion with
\begin{equation}
\mc H_0 = {1\over \sqrt{2}}\left(\matrix{0\cr v} \right),
\label{eq:Higg0}
\end{equation}
where $v$ is a constant.
%For the case of constant Higgs field,
In four dimensions, the classical equation of motion leads to the following relation
\begin{eqnarray}
R = {8 {\lambda} \over \beta} {\mc H_0}^\dagger {\mc H_0}
\label{eq:eomR}.
\end{eqnarray}

In the present model the Goldstone like mode, $S_{\rm G}$ is a linear combination of $h$ and the Higgs field.
We denote the full Higgs field multiplet as
\begin{equation}
\mc H = \mc H_0+ \hat \mc H,
\end{equation}
where we parametrize $\hat \mc H$ as
\begin{equation}
\hat\mc H = {1\over \sqrt{2}}\left(\matrix{\phi_1+i\phi_2\cr \phi_3+i\phi_4}
\right)
\end{equation}
and denote the real
Higgs field as $\Phi_3 = v + \phi_3$.

\subsection{Gauge Fixing}
We now consider the quadratic part of the Weyl meson Lagrangian including
its coupling to Higgs in four dimensions.
This will allow us to determine the gauge fixing Lagrangian. We expand
the Higgs field around its classical solution, given in Eqs. \ref{eq:Higg0}
and \ref{eq:eomR}. As in the case of the toy model, discussed in section 2,
we also need to include the graviton field, which gets mixed with the
Higgs and Weyl meson field. We redefine the graviton field as,
\begin{equation}
h_\alpha^\beta = {1\over 4} v\sqrt{\beta} h_\alpha^{\prime\beta},
\end{equation}
so that it's kinetic energy term is properly normalized.
Our Lagrangian is given by
\ba
\sqrt{-\bar g} \mc L'&=&\sqrt{-\bar g}{\beta\over 4}\mc H^\dagger\mc H \bar R' -\sqrt{-\bar g} \lambda
(\mc H^\dagger\mc H)^2 + \sqrt{-\bar g} {\bar g}^{\mu\nu}(D_\mu\mc H)^\dagger
D_\nu\mc H \nonumber \\
 &-& {1\over 4} \sqrt{-{\bar g}} {\bar g}^{\mu\rho}
 {\bar g}^{\nu\sigma}{\mc E}_{\mu\nu}{\mc E}_{\rho\sigma} + ....
\ea
Here we have explicitly displayed the contributions only due to the 
Higgs field, the Weyl meson and the graviton.
Expansion of above equation gives the following quadratic terms.
\ba
 \sqrt{-\bar g} \underline{\underline{\mc L}} &=& \sqrt{-g} \bigg\lbrace {\beta v^2\over 8}R_2+\sqrt{\beta} \left(\phi_3
+ {1\over \sqrt{\beta}} h\right)\left[ h^{\beta;\alpha}_{\beta;\alpha}-  h^{\beta;\alpha}_{\alpha;\ \beta}
- R^\alpha_\beta  h^\beta_\alpha\right] -\lambda v^2 \phi_3^2  \cr
&& +  {1\over 2}g^{\mu\nu}\partial_\mu\phi_i
\partial_\nu\phi_i +   {f^2v^2\over 2}\left(
1+{3\over 2}\beta\right)g^{\mu\nu} S_\mu S_\nu+ {3\beta\over 2}fv \left(\phi_3 + {1\over \sqrt{\beta}} h\right)S^\kappa_{;\kappa}  \cr
&&+ fv \phi_3S^\kappa_{;\kappa}
- {1\over 4}g^{\mu\rho}
g^{\nu\sigma}{\mc E}_{\mu\nu}{\mc E}_{\rho\sigma} + {4\lambda v^2\over \beta}\left(-{1\over 4} h^\alpha_\beta
 h^\beta_\alpha + {1\over 8}  h^2\right) \cr
&& \cr
&&+ g^{\mu\nu}{1\over 2}\biggl[-gv\left(\partial_\mu\phi_1 A_{\nu}^{2} +\partial_\mu\phi_2 A_{\nu}^{1}-\partial_\mu\phi_4 A_{\nu}^{3} \right)\cr
&&- g^{\prime} v B_{\mu}\partial_{\nu}\phi_4 + {g^2 v^2\over4 } A_{\mu}^{i} A_{\nu}^{i}  + {{g^{\prime}}^2 v^2\over4 }B_{\mu}B_{\nu}
- {g g{\prime}\over 2} B_{\mu} v^2 A_{\nu}^{3} \biggr]\bigg\rbrace. \label{eq:Expansion}
\ea
For consistency we should keep terms only at leading order in the coupling
$1/\sqrt{\beta}$. At higher orders in $1/\sqrt{\beta}$ we also need to
include the contribution due to graviton loops.

To eliminate the terms which mix $S_\kappa$ with $\phi_3$ and $ h$
we choose the following gauge fixing Lagrangian:
\begin{equation}
{\mc L}^{1}_{gf} = -{1\over 2\xi}\left[S^\kappa_{;\kappa}
+\xi M_3\phi_3 + \xi M_h h\right]^2,
\end{equation}
where
\ba
M_3 &=& vf(1+3\beta/2) \approx 3vf\beta/2,\cr
M_h &=& {3vf\over 2}\sqrt{\beta}.
\ea
The quadratic Lagrangian now takes the form
\begin{eqnarray}
\sqrt{-\bar g} \left(\underline{\underline{\mc L}}+\mc L^1_{gf}\right) &=&\sqrt{-g}
\bigg\lbrace{\beta v^2\over 8}R_2+\sqrt{\beta} \left(\phi_3
+ {1\over \sqrt{\beta}} h\right)\left[ h^{\beta;\alpha}_{\beta;\alpha}-  h^{\beta;\alpha}_{\alpha;\ \beta}
- R^\alpha_\beta  h^\beta_\alpha\right] \cr
&&-\lambda v^2 \phi_3^2 +  g^{\mu\nu}{1\over 2}\partial_\mu\phi_i
\partial_\nu\phi_i +  g^{\mu\nu} S_\mu S_\nu {f^2v^2\over 2}\left(
1+{3\over 2}\beta\right)\cr
&&+ {\lambda v^2\over 2\beta}\left(h^2-2 h^\alpha_\beta
 h^\beta_\alpha   \right) -{1\over 2\xi}  \left(S^\kappa_{;\kappa}\right)^2
-{\xi\over 2}\left(M_3\phi_3+ M_h h\right)^2 + \cdots\bigg\rbrace,
\end{eqnarray}
where we have not explicitly displayed the terms involving gauge fields other than
$S_\m$. In this model the Weyl meson mass  is found to be
\begin{equation}
M_S = vf\sqrt{1+3\beta/2} \approx \sqrt{3\beta/2}\, vf.\label{eq:Weylmass2}
\end{equation}

We still need to eliminate the mixing term which involves the scalar field
and the graviton. We notice that besides the term which mixes $\phi_3$ with
the derivative of the graviton field we also have mass terms which mix
$\phi_3$ with $h$. We define new fields $\tilde \phi_3$
and $\tilde h$,
\begin{eqnarray}
\phi_3 &=& \cos\theta\tilde\phi_3 - \sin\theta \tilde h,\nonumber\\
h &=& \sin\theta\tilde\phi_3 + \cos\theta \tilde h.
\label{eq:diagonalization}
\end{eqnarray}
We may rotate the entire graviton multiplet such that
\begin{equation}
h^\alpha_\beta = {1\over 4} g^\alpha_\beta\sin\theta\tilde\phi_3
+ \cos\theta \tilde h^\alpha_\beta.
\end{equation}
This leads to Eq. \ref{eq:diagonalization} for the trace part of the multiplet
$h$ with the traceless part remaining unchanged up to an overall rescaling
factor
\begin{equation}
h^\alpha_\beta - {1\over 4} g^\alpha_\beta h = \cos\theta
\left(\tilde h^\alpha_\beta - {1\over 4} g^\alpha_\beta \tilde h\right).
\end{equation}
The mixing angle $\theta$ that diagonalizes the
$\tilde\phi_3$ and $\tilde h$ mass matrix is given by
\begin{equation}
\tan \theta = {1\over \sqrt{\beta}} + \cdots ,
\end{equation}
where we have displayed only the dominant term.

We next choose another gauge fixing term so as to eliminate the term that
mixes the $\tilde \phi_3$ field with the graviton. At leading order the mixing
term may be written as
$$\sqrt{\beta} \tilde\phi_3\left(\tilde h^{;\alpha}_{;\alpha} - \tilde
h^{\beta;\alpha}_{\alpha;\beta}\right).$$
In order to eliminate this we use the
following gauge fixing term:
\begin{eqnarray}
\mc L^{2}_{g f}& = &{1\over 2\xi}\bigg[D^{\nu}\tilde h_{\mu\nu} - D_{\mu}\tilde h -
{\sqrt{\beta}\xi} D_{\mu}\tilde\phi_3 \bigg]
\bigg[D_{\sigma}\tilde h^{\mu\sigma} - D^{\mu}\tilde h -
{\sqrt{\beta}\xi} D^{\mu}\tilde\phi_3 \bigg]\nonumber \\
&=& {1\over 2\xi}\bigg[(D^{\nu}\tilde h_{\mu\nu} - D_{\mu}\tilde h)^2 +
{\beta\xi^{2}} D_{\mu}\tilde\phi_{3}D^{\mu}
\tilde\phi_3 \bigg]
+\sqrt\beta \tilde\phi_{3}\bigg[\tilde h^{\nu;\mu}_{\mu;\nu}- \tilde h^{\sigma;\mu}_{\sigma;\ \mu}\bigg].
\end{eqnarray}
In terms of the graviton field $h_{\alpha\beta}$, the quadratic terms in the
graviton field are given by,
\begin{eqnarray}
 {\mc L}_{gr}& = &{1\over2}\bigg[ h^{\beta}_{\alpha ;\nu}h^{\alpha ;\nu}_{\beta}
-  h_{ ;\nu}h^{ ;\nu}  + 2 h_{;\beta}h^{\beta ; \mu}_{\mu} \nonumber\\
 &&- 2h^{\nu ; \alpha}_{\beta }h^{\beta}_{\alpha;\nu}
+4 h^{\nu}_{\beta}h^{\beta} _{\alpha}R^{\alpha} _{\nu} -2 hh^{\nu} _{\beta} R^{\beta} _{\nu}
+ 2 \left({h^{2}\over 8} - {1\over 4} h^{\alpha}_{\beta} h^{\beta}_{\alpha}  \right)R  \bigg].
\end{eqnarray}
We next perform the transformation, Eq. \ref{eq:diagonalization}, and add
the gauge fixing term $L^{2}_{g f}$. At leading order,
we find the following quadratic terms involving the transformed
graviton field,
\begin{eqnarray}
{\mc L}_{gr} + \mc L^{2}_{g f} & =& {1\over 2}\tilde{h}^{\beta}_{\alpha ;\nu}\tilde{h}^{\alpha ;\nu}_{\beta}
+ \left({1\over 2} + {1\over 2\xi}\right)\tilde{h}_{ ;\nu}\tilde{h}^{ ;\nu}
- {1\over \xi}\tilde{h}_{;\beta}\tilde{h}^{\beta ; \mu}_{\mu}
 + \left(-1+{1\over 2\xi}\right){\tilde{h}^{\mu ;\nu}_{\nu}\tilde{h}^{\sigma}_{\mu ; \sigma}} + \cdots .
\label{eq:graviton}
\end{eqnarray}
Here we have dropped terms involving the background curvature since those
are higher order in $\lambda$ and negligible compared to the remaining terms.
For completeness, we obtain the expression for the graviton propagator
in this model in the Appendix.

The scalar field quadratic terms may be written as
\begin{equation}
\mc L_\phi = {1\over 2} (\beta\xi-1/2) \partial_\mu\tilde \phi_3 \partial^\mu\tilde
\phi_3 -{1\over 2} (2\lambda v^2 + \xi M_3^2)\tilde\phi_3^2.
\end{equation}
We should rescale the scalar field in order to obtain the canonical kinetic
energy term
\begin{equation}
\phi = \sqrt{\xi\beta-{1\over 2}}\,\tilde\phi_3 \approx \sqrt{\xi\beta}\,\tilde\phi_3.
\end{equation}
This gives
\begin{equation}
\mc L_\phi = {1\over 2} \partial_\mu \phi \partial^\mu
\phi -{1\over 2} M_\phi^2\phi^2,
\end{equation}
where
\begin{equation}
M_\phi\approx {3\over 2}\, vf\sqrt{\beta}.
\end{equation}
Hence we find that there does exist a physical scalar degree of
freedom in this model.
We so far do not know the value of the gauge parameter $f$. If we take
this to be of order unity then this has a mass much larger than the electroweak
mass scale. The couplings of this particle
to visible matter are suppressed
by powers of $1/\sqrt{\beta}$, hence these would be very small.
The Weyl meson acts as a dark matter in this model \cite{ChengPRL}. Its 
propagator is given in Eq. \ref{eq:Weyl_propagator} where its mass is now given in Eq. \ref{eq:Weylmass2}. 

The Higgs particle is normally desired in Standard Model since
it leads to perturbative unitarity \cite{Joglekar,Cornwall}. This means that if the Higgs
particle is absent then some of the amplitudes might grow rapidly
with energy. Our model does not appear to respect this.
Furthermore we also generate additional vertices in our model due
to the terms raised to fractional powers. Here we have chosen these regulator
terms to be proportional to the Higgs field raised to fractional powers.
Hence we obtain additional amplitudes, which grow with energy as powers
of $p^2/v^2$,
where $p^2$ is some momentum scale characteristic of the experiment
and $v$ is the electroweak scale. We essentially expect such amplitudes from
an infinite series of the vertices generated by these regulator terms
in the action. Hence the theory cannot be treated perturbatively at
energy scale beyond $v$. This implies that the theory is non-perturbative
or strongly coupled beyond the electroweak scale, as might be expected
for any theory which does not predict a Higgs particle in the physical
spectrum. It remains to be seen by experiments such as LHC if this
scenario is realized in nature.

\section{Locally Scale Invariant Standard Model with a Real Scalar Field}\label{sec:dilaton}
The main problem with the theory discussed in  
section \ref{sec:sism}
is that it is not perturbatively reliable
beyond the electroweak scale.
In this section we consider a generalization of the model, by adding
another scalar field, $\chi$, which is a singlet under the
electroweak symmetry transformations.  
 In this case we use the real scalar field, $\chi$, as the regulator scalar
field. The classical value of the field $\chi$ is taken to be much larger
than the electroweak mass scale. In this case
%we shall find that 
perturbation theory
breaks down only for energy scales much larger than the electroweak 
mass scale. 
The action given in Eq. \ref{eq:S_EW_d} now generalizes to
\ba
\mathcal{S} &=& \int d^dx \sqrt{-\bar g}\Bigg(\left[{\beta\over 8} \chi^2 +
{\beta_1\over 4} \mc H^\dag \mc H\right]\bar R'
+\bar g^{\mu\nu} (D_\mu \mc H)^\dag(D_\nu \mc H) + \frac12\bar g^{\mu\nu}
(D_\mu \chi) (D_\nu  \chi)\nonumber\\
 &&- {1\over 4}\lambda \chi^4 (\chi^2)^{\, -\delta} - {1\over 4}\lambda_1\left[
2\mc H^\dag \mc H - \lambda_2 \chi^2\right]^2 (\chi^2)^{\, -\delta}
\Bigg),
\label{eq:S_dilaton_d}
\ea
where we have set the regulator field
$\Phi=\chi$. The scalar potential of this type has been considered earlier in
Refs. \cite{JM,Shaposhnikov:2008a,Nishino2009}. In Eq. \ref{eq:S_dilaton_d}
we have not displayed the kinetic energy terms of the gauge fields
and the terms involving the fermions. These remain same as in Eq.
\ref{eq:S_EW_d}.  We shall choose the couplings $\beta$ 
and $\lambda_1$ to be of order unity. The couplings $\lambda,\lambda_2 \ll 1$,
such that $\lambda\sim H_0^2/M_{\rm PL}^2$ and $\lambda_2\sim v^2/M_{\rm PL}^2$.

The potential shown in Eq. \ref{eq:S_dilaton_d} appears fine
tuned and one might expect that quantum corrections will lead to acute
fine tuning problems in this case. In particular the couplings $\lambda$ and
$\lambda_2$ have to be chosen to be very small. 
However it has been shown in Ref.
\cite{Shaposhnikov:2008a} that the smallness of $\lambda_2$ does not cause
any problems. The potential is
stable due to the underlying scale invariance of the model.
The value of $\lambda$ has to be chosen to be very small
since it contributes directly to the cosmological constant.
Its contribution to Higgs physics is very small and hence it cannot
lead to any fine tuning in the Higgs potential, i.e. the term proportional
to $\lambda_1$. The only problem that
might arise is that the small coupling $\lambda$ might itself require
fine tuning. This is directly related to the fine tuning problem of the 
cosmological constant and we ignore this in the present paper. 

The minimum of this potential is obtained for $\chi=0$ and $\mc H=0$.
However as in the earlier case %of Eq. \ref{eq:S_EW_d} 
we do not seek the classical solution which minimizes the potential. 
Instead we seek a time
dependent solution, where the FRW scale parameter $a(t) = a_0 e^{H_0t}$
and the scalar fields take some constant values. We shall also use
the electroweak $SU(2)$ symmetry to orient the Higgs field such that
only $\phi_3$ is non-zero. The equations of motion for $\chi$ and $\phi_3$
may be written as
\ba
{\beta R\over 4} &=& \lambda\chi^2 - \lambda_1\lambda_2(\phi_3^2- \lambda_2\chi^2)\label{eq:chi}, \\
{\beta_1 R\over 4} &=&  \lambda_1(\phi_3^2- \lambda_2\chi^2)
\label{eq:phi3}
\ea
respectively. In these equations we have not displayed the derivative terms
involving the scalar fields, since we seek a solution where these fields
are constant. The Einstein's equations may be written as
\ba
{1\over 4} (\beta\chi^2 + \beta_1\phi_3^2)\left(R_{\mu\nu} - {1\over 2}R g_{\mu\nu}\right) = T_{\mu\nu}.
\ea
The $0-0$ component of this equation gives
\ba
3 (\beta\chi^2 + \beta_1\phi_3^2) \left({\dot a\over a}\right)^2
= \lambda\chi^4 + \lambda_1(\phi_3^2- \lambda_2\chi^2)^2.
\label{eq:Einstein_00}
\ea
From Eq. \ref{eq:chi} and \ref{eq:phi3} we find
\begin{eqnarray}
(\beta + \lambda_2\beta_1) {R\over 4} &=& \lambda\chi^2,\nonumber\\
\phi_3 &=& \zeta\chi,
\end{eqnarray}
where
\begin{equation}
\zeta = \left(\lambda_2+ {\lambda\over\beta\lambda_1\left({1\over \beta_1}
+{\lambda_2\over \beta}\right)}\right)^{1/2}.
\end{equation}

As mentioned before we seek a solution of the form $\chi = \chi_0 = {\rm constant} \approx M_{\rm PL}$ and
 $a(t) = a_0e^{H_0t}$ where the Hubble
parameter $H_0$ is constant. Since $R=12H_0^2$, we find
\begin{equation}
H_0^2 = {\lambda\chi_0^2\over 3(\beta + \beta_1\lambda_2)}.
\label{eq:classicalChi0}
\end{equation}
It is easy to verify that Eq. \ref{eq:Einstein_00} is consistent with this result.
Hence we find an acceptable classical solution. We next make a quantum
expansion around this solution, such that
\begin{eqnarray}
\chi &=& \chi_0 + \hat\chi,\nonumber\\
\phi_3 &=& \phi_{3,0} + \hat\phi_3.
\end{eqnarray}
In the present case we expect that the
Goldstone like mode, $S_{\rm G}$, that gets eaten by the Weyl meson
would dominantly be a linear combination of
$\hat\chi$ and graviton along with a small contribution from $\hat \phi_3$.

%%%%%%%%%%%%%%%%%%%%%%%%%%%%%%%%%%%%%%%%%
To obtain the proper normalization of the graviton field kinetic 
energy we set
\ba
h'^{\beta}_{\alpha} = {4\over {\sqrt{\beta \chi_0 ^2 + \beta_1 v^2}} }h^{\beta}_{\alpha}\,,
\ea
in analogy to Eq. \ref{eq:redefine}.
Quantum expansion of Eq. \ref{eq:S_dilaton_d} in four dimensions gives following quadratic terms,
\ba
 \sqrt{-\bar g} \underline{\underline{\mc L}} 
&=& \sqrt{-g}\bigg[{1\over 2}f^2 \left\{\chi_0^2 \left(1+ {3\beta\over 2}\right) + v^2 \left(1+ {3\beta_1\over 2}\right)\right\}S^{\mu}S_{\mu} 
+{1\over 2}D^{\mu}\phi_i D_{\mu}\phi_i +{1\over 2}D^{\mu}\hat\chi D_{\mu}\hat\chi\cr
&& + \left\{{\beta R\over 8}-{3\over 2}\chi_0 ^2 \left(\lambda + \lambda_1 \lambda_2 ^2\right) +{\lambda_1 \lambda_2\over 2}v^2 \right\}\hat\chi ^2
-\lambda_1 v^2 \phi_3 ^2+ {R_2\over 8}\left( \beta \chi_0 ^2 + \beta_1 v^2\right)\cr
&&+ {\beta \chi_0 \hat\chi + \beta_1 v \phi_3 \over \sqrt{\beta \chi_0 ^2 + \beta_1 v^2}}\left( h^{\beta;\alpha}_{\beta;\alpha}-  h^{\beta;\alpha}_{\alpha;\ \beta}
- R^\alpha_\beta  h^\beta_\alpha\right)  
+ h\left( h^{\beta;\alpha}_{\beta;\alpha}-  h^{\beta;\alpha}_{\alpha;\ \beta}
- R^\alpha_\beta  h^\beta_\alpha\right)\cr
&&+ f S^{\mu}_{; \mu}\left\{\left({3\beta\over 2} + 1\right)\chi_0 \hat\chi + \left({3\beta_1\over 2} + 1\right)v \phi_3 +\frac32{\sqrt{\beta \chi_0^2 + \beta_1 v^2} } \ h\right\} \cr
&&+ {4\over \beta \chi_0^2 + \beta_1 v^2} \left\{{\beta \chi_0^2 R \over 2} + \lambda_1 v^4 -\left(\lambda + \lambda_1 \lambda_2 ^2\right)\chi_0^4\right\}\left(-{1\over 4} h^\alpha_\beta
 h^\beta_\alpha + {1\over 8}  h^2\right) \cr
&&+ 2\lambda_1 \lambda_2 v \chi_0 \phi_3 \hat\chi + \cdots \bigg].
\ea
Here we have not shown the gauge field kinetic energy terms. 

\subsection{Gauge Fixing}
We may eliminate the mixing terms of $S_{\mu}$ with other fields 
by adding the following
gauge fixing term,
\ba
{\mc  L}^{3}_{gf} = -{1\over 2\xi}\left[S^\kappa_{;\kappa}
+\xi \left(C_1\hat\chi + C_2 \phi_3 + C_3 h \right)\right]^2,
\ea
where,
\ba
C_1 = f\left({3\beta\over 2}+1\right)\chi_0; \ \ C_2 =f\left({3\beta_1\over 2}+1\right)v; \ \ C_3 = \frac32f\sqrt{\beta \chi_0^2 + \beta_1v^2}.
\ea
Hence we find
\ba
 \sqrt{-\bar g} \underline{\underline{\mc L}} &=& \sqrt{-g}\Bigg[{1\over 2}D^{\mu}\phi_i D_{\mu}\phi_i +{1\over 2}D^{\mu}\hat\chi D_{\mu}\hat\ch
+ {\beta \chi_0 \hat\chi + \beta_1 v \phi_3 \over \sqrt{\beta \chi_0 ^2 + \beta_1 v^2}}\left( h^{\beta;\alpha}_{\beta;\alpha}-  h^{\beta;\alpha}_{\alpha;\beta}
\right) + R'_2  \cr
&&- {1\over 2 \xi}\left( S^{\mu}_{;\mu} \right)^2+ {1\over 2}f^2 \left\{\chi_0^2 \left(1+ {3\beta\over 2}\right) + v^2 \left(1+ {3\beta_1\over 2}\right)\right\}S^{\mu}S_{\mu}
 \cr
&&+   \bigg\{{\beta R\over 8}-{3\over 2}\chi_0 ^2 \left(\lambda + \lambda_1 \lambda_2 ^2\right) -{\lambda_1 \lambda_2\over 2}v^2
-{\xi\over 2}C_1^2 \bigg\} \hat\chi ^2 - \left(\lambda_1 v^2 +{\xi\over2}C_2^2 \right)\phi_3 ^2 \cr
&&+  {\left(h^2-2 h^\alpha_\beta
 h^\beta_\alpha \right)\over 8\left(\beta \chi_0^2 + \beta_1 v^2\right)} \left\{-{R\over2}\left({\beta_1 v^2}\right) + \lambda_1 v^4 -\left(\lambda + \lambda_1 \lambda_2 ^2\right)\chi_0^4\right\} \cr
&&- {\xi\over2}C_3^2 \ h^2 + \left(2\lambda_1 \lambda_2 v \chi_0 -\xi C_1C_2\right)\phi_3 \hat\chi
-\left({\beta \chi_0 R \over 4 \sqrt{\beta \chi_0 ^2 + \beta_1 v^2}} +\xi C_1C_3\right)\hat\chi h \cr
&&-\left({\beta_1 v R \over 4 \sqrt{\beta \chi_0 ^2 + \beta_1 v^2}} +\xi C_2C_3\right)\phi_3 h +\cdots \Bigg],\ea
where $R_2'$ is kinetic energy terms of gravitons.

The mass of the Weyl meson in this theory is given by
\begin{equation}
M_S^2 = f^2\left[\chi_0^2\left (1+ {3\beta\over 2}\right) + v^2 \left(1+ {3\beta_1\over 2}\right)\right].
\end{equation}
If we choose parameters such that $\beta_1 v^2\ll \beta \chi_0^2$, this is 
essentially the same as that found in the case of a single scalar field in
section 2. The scalar field mass term may be written as
$-\Phi^T M^2 \Phi/2$,
where 
\ba
\Phi =\left(\matrix{\hat{\chi}\cr \phi_3 \cr h} \right). 
\ea
The  mass matrix of scalar field may be decomposed as
\begin{equation}
M^2 = M_0^2 + \Delta M^2,
\end{equation}
where the unperturbed mass matrix $M_0^2$ is taken to be
\begin{equation}
M_0^2 = \left(\matrix{a^2 & ab & ac\cr ab & b^2 + \epsilon_2 & bc\cr
ac & bc & c^2}  \right).
\end{equation}
The perturbation term is given by
\begin{equation}
\Delta M^2 = \left(\matrix{\epsilon_1 & \epsilon_4 & \epsilon_5\cr \epsilon_4 & 0
& \epsilon_6\cr
\epsilon_5 & \epsilon_6 & \epsilon_3}  \right).
\end{equation}
Here $a^2 = \xi C_1^2$, $b^2 = \xi C_2^2$, $c^2 = \xi C_3^2$ and
$\epsilon_2 = 2\lambda_1 v^2$.
We point out that $a^2$, $c^2$ and $ac$ are of order $\chi_0^2$,
$ab$ and $bc$ of order $\sqrt{\lambda_2}\chi_0^2$,
$b^2$ and
$\epsilon_2$ of order $\lambda_2\chi_0^2$, $\epsilon_4$ of
order $\lambda_2^{3/2}\chi_0^2$ and the remaining
parameters of order $\lambda_2^2\chi_0^2$ or $\lambda\chi_0^2$. In the unperturbed term it
is useful to keep the higher order terms $ab$, $bc$, $b^2$ and $\epsilon_2$
since it allows us to properly identify the Higgs particle, with
gauge invariant mass. If we keep only the dominant terms in the unperturbed
matrix then at leading order the Higgs particle is massless and at next
to leading order its mass is gauge dependent.

We next diagonalize the unperturbed part of the mass matrix. Here it is
convenient to expand the unperturbed eigenvalues and eigenvectors
in powers of $\lambda_2$. As we compute the higher order corrections
due to the perturbation $\Delta M^2$ we also need to simultaneously
correct the unperturbed solution to the accuracy of the calculation.
At leading order the eigenvalues are given by
\begin{equation}
m_1^2 \approx a^2+b^2+c^2 = \xi(C_1^2+C_2^2+C_3^2),\ \ m_2^2\approx \epsilon_2=2\lambda_1 v^2, \ \ m_3^2= 0.
\end{equation}
The corresponding eigenvectors are
\begin{equation}
{1\over N_1}\left(\matrix{C_1\cr C_2\cr C_3}\right),\ \
{1\over N_2}\left(\matrix{-C_1C_2\cr (C_1^2+C_3^2)\cr -C_2C_3}\right),\ \
{1\over N_3}\left(\matrix{-C_3\cr 0\cr C_1}\right).
\end{equation}
Here $N_1=\sqrt{C_1^2 + C_2^2+C_3^2}$, $N_2=\sqrt{(C_1^2+C_2^2+C_3^2)(C_1^2+C_3^2)}$ and $N_3=\sqrt{C_1^2+C_3^2}$ are normalization factors.

We identify the particle of mass $m_1$ as the Goldstone type mode that
generates the mass of the
Weyl meson. The mass of this mode depends on the gauge parameter $\xi$ and
is of the order of Planck mass for $\xi$ of order unity. The particle of
mass $m_2$ is identified with the Higgs and the third particle as one
of the components of the graviton.
We, therefore, find that the longitudinal component of the Weyl meson, $S_{\rm G}$,
is dominantly a linear combination of $\hat\chi$ and the graviton field
$h$ along with a very small contribution, of order $\beta_1 v/M_{\rm PL}$, from
$\phi_3$. Here we have assumed that $\beta_1\gg 1$.  

\subsection{Phenomenological Implications}
The Higgs particle is present in the physical spectrum in this
model in contrast to the model 
with pure Higgs field \cite{ChengPRL}, 
discussed in section \ref{sec:sism}. Since the Higgs particle
couples directly to the Weyl meson there might be some
possibility of detecting the Weyl meson at the present or future colliders.
The Weyl meson is also a candidate for dark matter \cite{ChengPRL}. 
In the present case it's cosmological implications may differ 
significantly from those of the model with pure Higgs field. 

The coupling of the Higgs and the Weyl meson can be obtained from the
kinetic energy term of the Higgs and the term proportional to $\beta_1$.
Here we shall work in the unitary gauge and consider only the physical
particles. The relevant terms are
\begin{equation}
g^{\mu\nu} (D_\mu \mc H)^\dag(D_\nu \mc H) = 
-f g^{\mu\nu}S_\nu\phi_3\partial_\mu\phi_3 + vf^2g^{\mu\nu}S_\mu S_\nu\phi_3
+ {1\over 2} g^{\mu\nu} f^2S_\mu S_\nu\phi_3^2 + \cdots 
\label{eq:interactions1}
\end{equation}
and
\begin{equation}
{\beta_1\over 4} {\mc H}^\dag {\mc H} R' = 
{3\beta_1f\over 4}\left[\phi_3^2 S^\mu_{;\mu} + 2vf\phi_3 S^\mu S_\mu
+ f\phi_3^2 S^\mu S_\mu\right] + \cdots
\label{eq:interactions2}
\end{equation}
We need to express the scalar field $\phi_3$ in terms of the physical 
scalar fields. However to a good approximation $\phi_3$ is simply the
physical Higgs field. Hence the interaction terms are directly
given by Eqs. \ref{eq:interactions1}-\ref{eq:interactions2}.

Let's first assume that the parameter $f$ is of order unity. 
In this case the Weyl meson mass is of order of the Planck mass
and it will give negligible contribution at the current particle colliders.  
However we may consider the case where $f$ and hence the Weyl meson
mass is sufficiently small that it can be produced in colliders such 
as LHC. In this case the coupling terms would become very small 
and hence again the probability to produce this particle at current colliders
is vanishingly small. However there does exist the possibility that
the parameter $\beta_1$ is sufficiently large, such that the product $\beta_1 f$
is not very small. In this case the first term on the right hand side
of Eq. \ref{eq:interactions2} may lead to an observable signal 
of the Weyl meson at the current and future colliders. Among the standard
model particles, the Weyl meson
couples direcly only to the Higgs meson.
Hence the Weyl meson may contribute to processes which produce a 
Higgs particle. We postpone a detailed
phenomenological investigation to future research. 

\section{Conclusions}\label{sec:conclusion}
Local scale invariance and the corresponding Weyl 
vector meson were first proposed in 1929 \cite{Weyl:1929}. It was later
found to be a candidate for dark matter \cite{ChengPRL} since it does
not couple directly to any of the Standard Model fields, except the Higgs.
In the present paper we have discussed the physics of the Weyl meson
in detail. In the case of the model proposed in Ref. \cite{ChengPRL},
the Higgs field is absent from the particle spectrum making the model
perturbatively unreliable beyond the electroweak scale. We have
discussed a generalization
of this model which includes an additional scalar field. In this case
the Higgs particle is present in the physical spectrum. Furthermore
we find that for a certain range of parameters the Weyl meson may
produce an observable signal in current or future colliders. If this parameter 
range is realized in nature, the Weyl meson may contribute in processes
which lead to the production of the Higgs particle.
Detailed phenomenological study of its signal is postponed for future
research. 

\section{Appendix}
In this Appendix we obtain the expression for the graviton propagator.
The quadratic terms in the Lagrangian, Eq. \ref{eq:graviton},
involving graviton field may be written as
\begin{eqnarray}
{\mc L}_{gr} + {\mc L}_{gf}^2
 = {1\over 2}\tilde {h}_{\alpha \beta}\bigg[D^{\alpha \beta , \mu \nu}\bigg]\tilde {h}_{\mu \nu}.
 \end{eqnarray}
Let us denote the graviton propagator by $P_{\mu\nu,\gamma\delta}$. At
leading order the propagator can be written as
\begin{eqnarray}
iP_{\mu\nu,\gamma\delta}  &=& {2i\over {k^2 + i\epsilon} } \bigg[ A({g_{\mu\gamma}g_{\nu\delta}
 + g_{\mu\delta}g_{\nu\gamma} }) + B(g_{\mu\nu}g_{\gamma\delta})
+ C{g_{\mu\nu}k_{\gamma}k_{\delta} + g_{\gamma\delta}k_{\mu}k_{\nu}\over k^{2} }\nonumber \\
&&+ D{g_{\nu\delta}k_{\mu}k_{\gamma}
  + g_{\nu\gamma}k_{\mu}k_{\delta} + g_{\mu\delta}k_{\nu}k_{\gamma} + g_{\mu\gamma}k_{\nu}k_{\delta}\over k^2 }
  + E {{k_\mu k_\nu k_\gamma k_\delta}\over k^4}
  \bigg],
\end{eqnarray}
where $k$ is the momentum carried by the graviton. To determine A, B, C, D and E we use the following relation:% the are functions of $k^2$,
\ba
 D^{\alpha\beta,\mu\nu}{P_{\mu\nu,\gamma\delta}} & = & {1\over 2}(g^{\alpha}_{\gamma}g^{\beta}_{\delta} + g^{\alpha}_{\delta}g^{\beta}_{\gamma}),
\ea
which implies
\ba
A = {1\over4},  \quad B = -{1\over6}, \quad C = {1\over3},\quad
D = -{1\over4}+{\xi\over 2},\quad E = -{2\over3}-{2\xi}-{1\over 2\xi}.
\end{eqnarray}
In the graviton propagator we have dropped terms proportional to the classical
curvature $R$ since they are higher order in the
coupling $\lambda$.

\bigskip
\noindent
{\large\sf\bfseries Acknowledgements}  \\
Naveen Kumar Singh thanks the Council
of Scientific and Industrial Research(CSIR), India for providing his Ph.D.
fellowships. His fellowship numbers is  F.No.09/092(0437)/2005-EMR-I.

\end{spacing}
\begin{spacing}{1}
\begin{small}

\end{small}
\end{spacing}
\end{document}